\documentclass{phc-proc4-auth}
\usepackage[norsk,english]{babel}
\usepackage{graphicx}
\usepackage{pstricks}
\usepackage{pst-plot}
\usepackage{epsf}
\usepackage{calc}
\journal{Physica C}
\begin{document}

\begin{frontmatter}
\title{Single vortices observed as they enter NbSe$_2$}
\author{{\AA}. A. F. Olsen\corauthref{cor}},
\corauth[cor]{Corresponding author.}
\ead{a.a.f.olsen@fys.uio.no}
\author{H. Hauglin},
\author{T. H. Johansen},
\author{P. E. Goa} and
\author{D. Shantsev}
\address{Dep. of Physics, University of Oslo, PO BOX 1048 Blindern, N-0316 Oslo, Norway}
\begin{abstract}
We observe single vortices as they penetrate the edge of a superconductor using 
a high-sensitivity magneto-optical microscope. 
The vortices leap across a gap near the edge, 
a distance that decreases with increasing applied field and sample thickness.
This behaviour can be explained by the combined effect of 
the geometrical barrier and bulk pinning.   
\end{abstract}
\end{frontmatter}

\section{Introduction}

As vortices enter a superconductor, their motion is determined by the
complex interplay between several competing forces: the Lorentz force from
the shielding currents tends to drag them into the sample, pinning
forces resist their motion, and finally they must overcome an edge barrier (the Bean-Livingston \cite{bean:1964} barrier
and the geometrical barrier \cite{zeldov:1994a}). 

The resulting behaviour has been studied both theoretically and
experimentally. It has been found that the forces near the edges 
can be a source of magnetic irreversibility even in the
absence of bulk pinning \cite{zeldov:1994a}, 
and can control the transport current distribution (\cite{paltiel:1998}, \cite%
{paltiel:2000}).


In this paper we report direct magneto-optical (MO)  observations 
of individual vortex motion near a superconductor's edge.

\section{Method}

\suppressfloats

We have used a magneto-optical microscope capable of resolving single
vortices to look at the detailed flux dynamics near the edge
for two samples of NbSe$_{2}$. The basic principle of the method 
is to let polarised
light pass through a magneto-optical indicator film placed close to the
sample surface, and then detect changes in the polarisation. 
The instrument is further described in \cite{goa:2003}. For
single vortex resolution the MO film must be very close to the sample, 
because at larger distances the
single vortex field quickly becomes smeared out. 

The field penetration experiments were performed on two NbSe$_{2}$ single
crystals \cite{Oglesby:1994}  with thickness 100 $\mu $m and  10 $\mu $m. The
sample shape was approximately a  rectangle with the smallest dimension 
$\approx$1 mm. The samples were initially cooled to 5 K in a residual
field of 0.05 mT. After cooling, an external
field $H_{a}$ was applied perpendicular to the plane of the sample and
ramped from 0 to 1 mT.  

\section{Observations}

\begin{figure}[t]
\par
\scalebox{0.5}{
\psset{unit=\textwidth}
\psset{origin={0,0.5}}
\pspicture(0,0.75)(1,1.05)
\rput[bl](0.05,0.75){
  \scaleboxto(0.29,0){
    \epsfbox{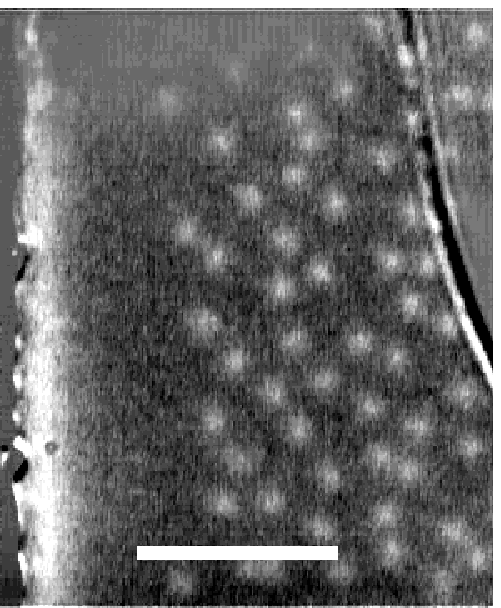}
  }
}
\rput[bl](0.35,0.75){
  \scaleboxto(0.29,0){
    \epsfbox{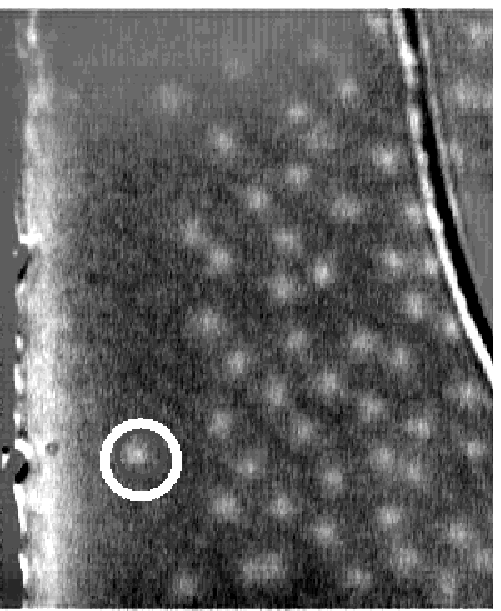}
  }
}
\rput[bl](0.65,0.75){
  \scaleboxto(0.29,0){
    \epsfbox{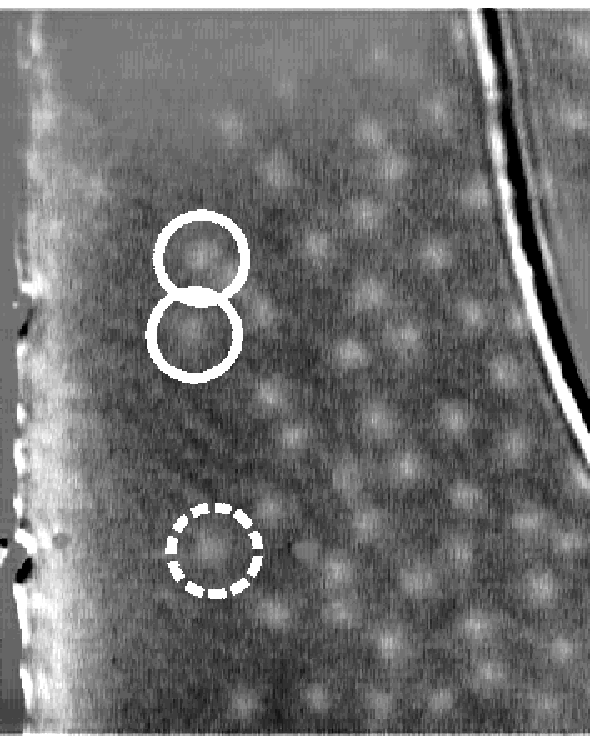}
  }
}
\endpspicture
}
\caption{MO images showing vortex penetration in a 100 $\protect\mu $m thick NbSe$_{2}$
crystal. The applied
field is $\approx 0.2$ mT and increases slightly from left to right.
In the second image a
vortex has appeared, marked with a white circle. In the third two more have
appeared, while the first one has moved slightly.
A wide vortex free band is seen between the edge (a narrow bright band on the left)
and the vortex-filled region. 
The scalebar is 10 $\protect%
\mu m.$}
\label{fig:tykk}
\end{figure}

\begin{figure}[tbp]
\par
\scalebox{0.5}{
\psset{unit=\textwidth}
\pspicture(0,0.65)(1,1.05)
\rput[bl](0.05,0.8){
  \scaleboxto(0.28,0){
    \epsfbox{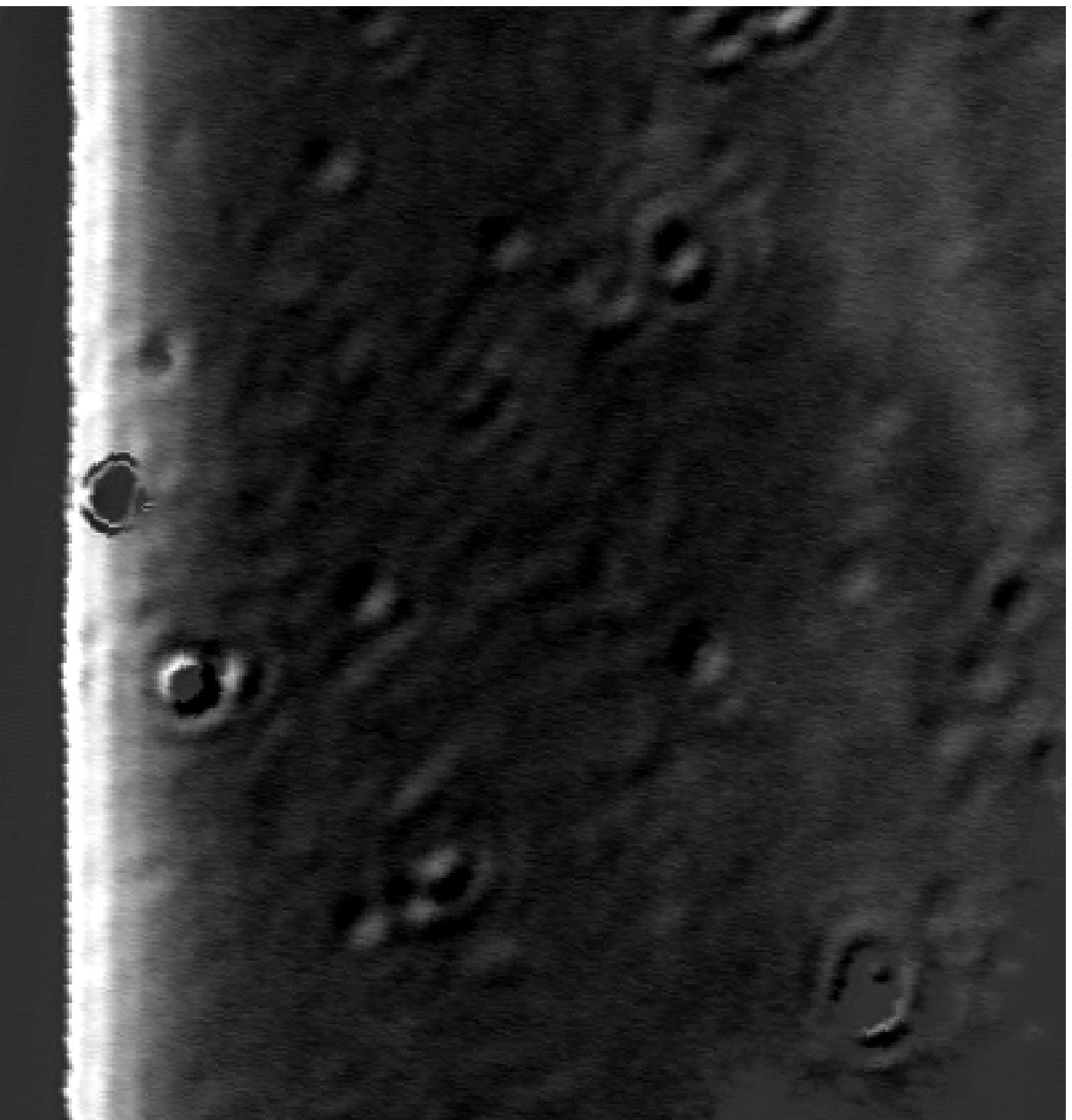}
  }
}
\rput[bl](0.35,0.8){
  \scaleboxto(0.28,0){
    \epsfbox{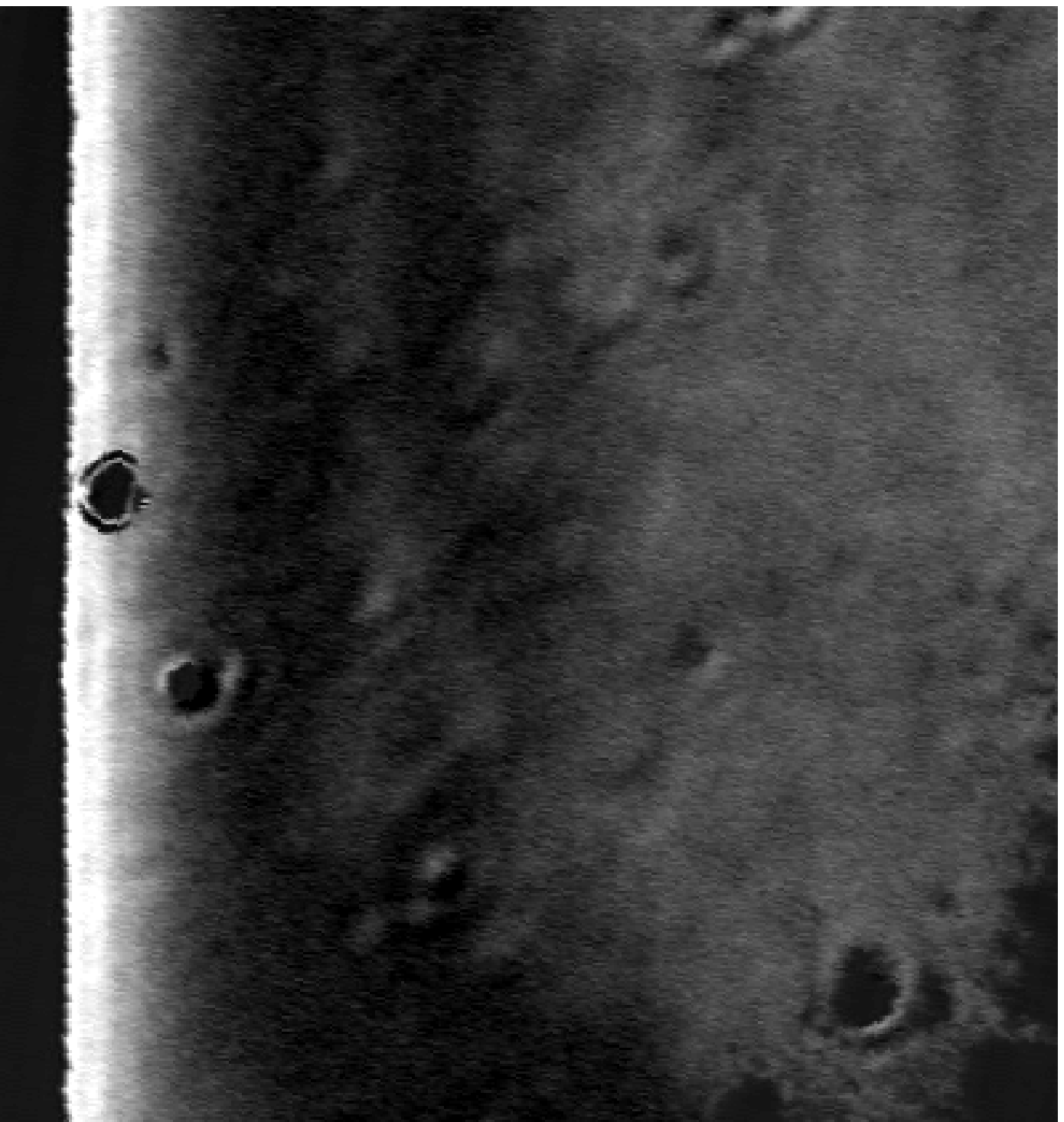}
  }
}
\rput[bl](0.65,0.8){
  \scaleboxto(0.28,0){
    \epsfbox{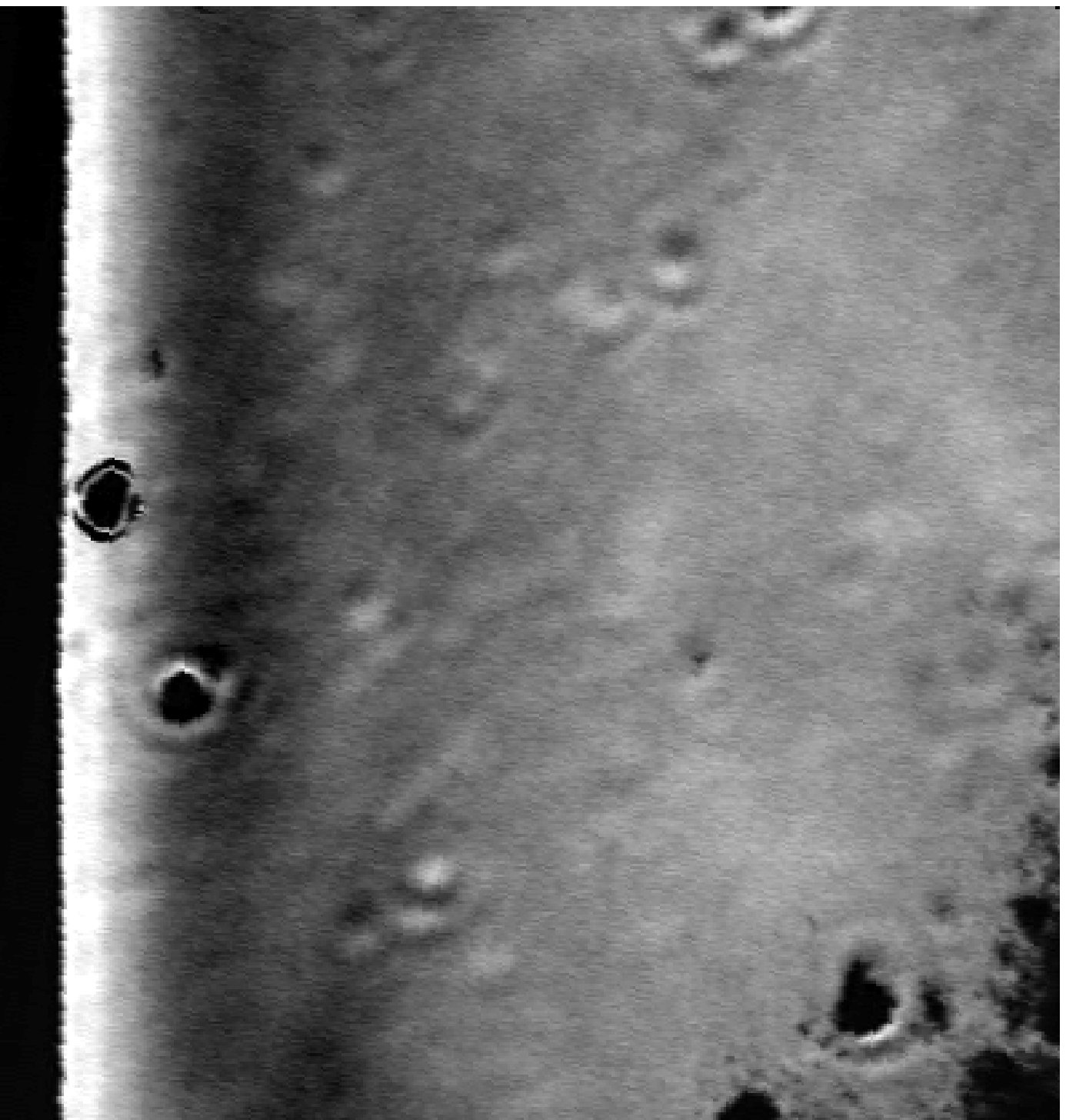}
  }
}

\rput[bl](0.05,0.62){
  \scaleboxto(0.28,0){
    \input{gjennomsnitt_3_43.tex}
  }
}

\rput[bl](0.35,0.62){
  \scaleboxto(0.28,0){
    \input{gjennomsnitt_250_260.tex}
  }
}

\rput[bl](0.65,0.62){
  \scaleboxto(0.28,0){
    \input{gjennomsnitt_524_544.tex}
  }
}

\endpspicture
}
\caption{MO images showing vortex penetration in a 10 $\protect\mu $m thick 
NbSe$_{2}$ crystal
at applied fields of 0.1, 0.5 and 1.0 mT (left to right). 
The sample edge is seen as a bright line on the left.
As the field increases, the flux front is moving towards the edge, and the flux free
zone near the edge shrinks.  
The graphs underneath show
profiles of flux density obtained from the images by averaging gray level in
each pixel column.}
\label{fig:tynn}
\end{figure}

Figure \ref{fig:tykk} shows snapshots of  vortex penetration into the 100 $%
\mu $m thick sample. Individual vortices are clearly seen as they enter
the sample. But we also see that near the edge of the sample a vortex free
band exists, approximately 5 $\mathrm{\mu m}$ wide. When a vortex enters the sample, 
it jumps across the band and gets pinned.
When
observing the vortex entry over a wider field range, we see that the
band remains (albeit thinning a bit as the field increases) as more vortices
penetrate, while the vortex density increases beyond the band. 
A full movie of vortex penetration during the field ramp can be found at
http://www.fys.uio.no/super/results/sv.


Figure \ref{fig:tynn} shows field distributions for the thin sample. 
Because of larger distance between the sample and the MO film,
individual flux quanta cannot be seen here. 
However, the gray levels still represent
the actual flux density. 
We have plotted profiles of flux density underneath the images
by averaging the gray-levels column-by-column.
For this sample we also see a vortex
free band near the sample edge.
The band width is again field dependent, but
it is much larger than for the thick sample at the same $H_a$.

\section{Discussion}

Both samples show a vortex free band near the edge. When vortices enter the sample,
they move quickly through this band and get pinned some distance $x$
from the edge. The three main observations to be explained are: (i) the band is shrinking
as the field increases, (ii) the band is wider for the thinner sample,
(iii) there is a narrow bright band at the edge. 
All these observations can be explained by  
the combined effect of the geometrical barrier and bulk pinning \cite{zeldov:1994a},
\cite{elistratov:2002}. A physical picture of the vortex entry process is the following: there is an energy cost associated with vortex formation at a superconductor's edge.
Once the barrier due to this energy is overcome, the vortex is pushed inwards by
the Meissner current. The Meissner force decays as the vortex gets deeper 
so at some point it is balanced by the pinning force, and the vortex stops.
The pinning force is larger for the thicker sample, so qualitatively this is consistent with our observation (ii) above.

However, this simple picture can not account for the first observation, and hence a more careful analysis is required. When vortex entry is governed by both bulk pinning and the geometrical barrier, the width of the vortex free band can be found from
the equation 
$
H_a/H_0+ d j_c/H_0 \ln{w/x} = 1 / \sqrt{x/w} 
$
obtained from Eq.(15) of Ref.\cite{elistratov:2002} in the limit $x \ll w$, where
$2w$ is the sample width, $d$ - thickness,
$j_c$ is the bulk critical current density, and $H_0$ is a normalization parameter.
The band width dependence $x(H_a,d)$ following from  
this equation is in qualitative agreement with both our observations (i) and (ii).
Besides, it predicts a stronger dependence of $x$ on $H_a$ for
a thinner sample, which is also the case experimentally. 

Finally, the bright band of high flux density near the sample
edge can be explained by a build up of partially penetrated 
vortices ``climbing'' the geometrical barrier.
In fact, individual vortices constituting this band can be resolved at low fields.
However, the width of this barrier is thought to be of the order of the sample 
thickness \cite{zeldov:1994a}, while  we
observe a much lower width of the band ($\approx 3 \mu $m) that is the same for both
samples.

\bibliographystyle{elsart-num}
\bibliography{../bib/bibliography}

\begin{thebibliography}{1}
\expandafter\ifx\csname url\endcsname\relax
  \def\url#1{\texttt{#1}}\fi
\expandafter\ifx\csname urlprefix\endcsname\relax\def\urlprefix{URL }\fi

\bibitem{bean:1964}
C.~P. Bean, J.~D. Livingstone, Phys. Rev. Lett. 12~(1) (1964) 14--16.

\bibitem{zeldov:1994a}
E.~Zeldov, A.~I. Larkin, V.~B. Geshkenbein, M.~Konczykowski, D.~Majer,
  B.~Khaykovich, V.~M. Vinokur, Phys. Rev. Lett. 73~(10) (1994) 1428--1431.

\bibitem{paltiel:1998}
Y.~Paltiel, D.~T. Fuchs, E.~Zeldov, Y.~N. Myasoedov, H.~Shtrikman, M.~L.
  Rappaport, E.~Y. Andrei, Phys. Rev. B 58~(22) (1998) 14763--14766.

\bibitem{paltiel:2000}
Y.~Paltiel, E.~Zeldov, Y.~N. Myasoedev, H.~Shtrikman, S.~Bhattacharya, M.~J.
  Higgins, Z.~L. Xiao, E.~Y. Andrei, P.~L. Gammel, D.~J. Bishop, Nature 403
  (2000) 398--401.

\bibitem{goa:2003}
P.~E. Goa, H.~Hauglin, {\AA}.~A.~F. Olsen, M.~Baziljevich, T.~H. Johansen, Rev.
  Sci. Ins. 74~(1) (2003) 141--146.

\bibitem{Oglesby:1994}
C.~S. Oglesby, E.~Bucher, C.~Kloc, H.~Hohl, J. Crys. Growth 137 (1994)
  289--294.

\bibitem{elistratov:2002}
A.~A. Elistratov, D.~Y. Vodolazov, I.~L. Maksimov, J.~R. Clem, Phys. Rev. B 66
  (2002) 220506.

\end{thebibliography}

\end{document}